\documentstyle[multicol,aps, prl]{revtex}

\date{\today}
\draft
\tighten
\begin{document}
\def\sqr#1#2{{\vcenter{\hrule height.3pt
      \hbox{\vrule width.3pt height#2pt  \kern#1pt
         \vrule width.3pt}  \hrule height.3pt}}}
\def\square{\mathchoice{\sqr67\,}{\sqr67\,}\sqr{3}{3.5}\sqr{3}{3.5}}
\def\today{\ifcase\month\or
  January\or February\or March\or April\or May\or June\or July\or
  August\or September\or October\or November\or December\fi
  \space\number\day, \number\year}

\def\Bbb{\bf}
\newcommand{\ww}{\mbox{\tiny $\wedge$}}
\newcommand{\pp}{\partial}
\def\d{{\nabla}}

\title{Absence of a VVDZ Discontinuity in $AdS_{AdS}$}
\author{Andreas Karch, Emanuel Katz and Lisa Randall}

\address {\qquad \\ Center for Theoretical Physics \&
Department of Physics\\
Massachusetts Institute of Technology\\
77 Massachusetts Avenue\\
Cambridge, MA 02139, USA
}

\maketitle

\begin{abstract}
We clarify the role of gauge invariance for the theory of an
$AdS_4$ brane
embedded in $AdS_5$. The presence of a
nonvanishing
mass parameter even for the lightest KK mode of the graviton
indicates that all of the spin-2 modes propagate
five polarization states. Despite this fact, it was
shown earlier that the classical theory has a smooth limit as the mass
parameter is taken to zero. We argue that locality in the fifth
dimension ensures that this property survives at the quantum level.
\end{abstract}


\begin{multicols} {2}

\section{Introduction}

Recently it was observed that a four dimensional $AdS$ theory
derived from a brane embedded in $AdS_5$ reproduces correctly a
four-dimensional theory of gravity, despite the presence of a mass parameter
\cite{kr}. It was shown by M. Porrati \cite{mp} and I. Kogan
et.al. \cite{oxford} that this is not in contradiction with the
Van Dam Veltman Zakharov (VVDZ) \cite{vvdz} discontinuity, which
in flat space says that the limit of a massive graviton as the mass
goes to zero is not a
massless graviton, because additional polarization states
survive. However, it was observed in \cite{duff} that
the VVDZ discontinuity could reappear for
a single massive spin 2 excitation at the one-loop level.

In this paper we want to clarify
the counting of degrees of freedom and
the role of gauge invariance in the
$AdS_{AdS}$
setup\footnote{We denote the theory
of \cite{kr} in which an $AdS_4$ brane is embedded inside
$_{AdS_5}$ as $AdS_{AdS}$.}.
In addition, we will argue on the basis of locality
of the higher dimensional theory that there cannot exist a VVDZ
discontinuity, even on the quantum level.

Our outline is as follows.
In the next section, we summarize the counting of degrees of freedom.
In Section 3, we discuss 
the fluctuation spectrum around an $AdS_4$ brane to show
explicitly that the only propagating modes of the graviton are the
transverse traceless modes. 
We speculate
about the possibility of a non-local 4D effective description
in Section 4.
In Section 5 we 
will discuss possible quantum effects and will argue on 
the basis of locality in the fifth dimension, that no
VVDZ discontinuity can appear even at the quantum level.

\section{Counting Propagating Degrees of Freedom}

The theory we consider is fundamentally a five-dimensional theory
of gravity. Therefore, there are naively fifteen polarizations.
It is clear that one can go into axial gauge, where there are only
ten polarization states. We can now proceed in several different ways
to further eliminate gauge artifacts. We can
first choose Transverse-Traceless (TT) gauge, in which both the longitudinal
states and the trace are eliminated
for all the KK modes. This
explicitly demonstrates the absence of a physical ghost scalar.
After making this gauge
choice, there are still three residual gauge invariances.
However these gauge transformations do not
fall off at the boundary of AdS. 
This means we can not eliminate normalizable
fluctuations with these non-normalizable gauge transformations.
If we expand the graviton into normalizable modes and a non-normalizable
vector field as in \cite{mp2}, only the non-normalizable
St\"uckelberg like vector field can be set to zero.
We therefore remain with a tower of normalizable massive spin 2 modes,
each with 5 degrees of freedom.
So just from counting degrees of
freedom, for our system
the absence of the VVDZ discontinuity still looks puzzling.
On the classical level the results of \cite{mp,kogan}
show that despite the mismatch in polarization states
there is no discontinuity in the propagator.
Without having gauge invariance to actually remove three
polarizations, one would generically expect that
the discontinuity would reappear when considering loop
corrections.
In Section 4 we will argue that it is really locality in
the fifth dimension which guarantees absence of the VVDZ discontinuity
even at the quantum level.

\section{Mode Analysis for the $AdS_4$ Brane}

In this section, we explicitly analyze the modes of the $AdS_4$
brane, considered in Ref. \cite{kr,matthew,andre,kogan} and
find that there is no ghost in the spectrum;
that is, the
TT modes suffice and any apparent ghost is a gauge artifact \cite{dvali}.
We are using the
same conventions and background solution as in \cite{kr}. 
The gauge transformations read
\begin{eqnarray}
\label{gaugetrafosfull}
h_{55} &\rightarrow & \quad h_{55} - 2 e^{-2A} \xi^{5'} \\
\label{munu}
h_{\mu \nu} &\rightarrow & \quad h_{\mu \nu} + (\d_{\mu} \xi_{\nu} +
\d_{\mu}
 \xi_{\nu}) +
2 A' g_{\mu \nu} \xi^5 \\
h_{\mu5} &\rightarrow & \quad h_{\mu5} - e^{-2A} \d_{\mu} \xi^5 +
g_{\mu \nu} \xi^{\nu '}
\end{eqnarray}

It is clear there is a choice of gauge
where $h_{\mu 5} = h_{55}=0$. We want to study transformations
that leave this gauge choice invariant. They are required to
satisfy
\begin{eqnarray}
\label{residual}
\xi_5 &=& \quad \epsilon^5(x) \\
\label{residualmunu}
\xi_{\mu} &=& \quad G \d_{\mu} \epsilon^5 + \epsilon_{\mu}(x)
\end{eqnarray}
with $G = \int_0^r e^{-2A(\tilde{r})} d\tilde{r}$.

The equations of motion demand
\begin{eqnarray}
\label{fullfluctuations}
55: & \quad &(e^{2A} h')' =0 \\
5\mu: & \quad &-\frac{1}{2} \d_{\mu} h' + \frac{1}{2} \d^{\rho}
h_{\mu
\rho}' =0 \\
\label{fullfluctuationsij}
\mu \nu: & \quad &e^{2A} (\frac{1}{2} h_{\mu \nu}'' + 2 A' h_{\mu
\nu}') -
\frac{1}{2} \d^2 h_{\mu \nu} + \\ \nonumber
&\quad &  \frac{1}{2} g_{\mu \nu} e^{2A} A' h' -
    \frac{1}{2} \d_{\mu} \d_{\nu} h + \\ \nonumber
& \quad &
\frac{1}{2} ( \d^{\rho} \d_{\mu} h_{\nu \rho} + \d^{\rho} \d_{\nu}
h_{\mu \rho})+ 3 \Lambda h_{\mu \nu} =0 \ .
\end{eqnarray}
with the boundary condition $h'_{\mu \nu}=0$

A generic solution to the bulk equations of motion can be brought
into the form \cite{kr}
\begin{equation}
\label{hsolved}
 h_{\mu \nu} = h^{TT}_{\mu \nu} + 2 G \; (\d_{\mu} \d_{\nu} -
\Lambda g_{\mu \nu}) \Phi - \frac{2}{3} \lambda g_{\mu \nu} \Phi
\end{equation}
with $\Phi$ satisfying $\d^2 \Phi = 4 \Lambda \Phi$, where we have
used the $\epsilon_{\mu}$ degrees of freedom and the equations of
motion to fix $G(0)=0$.  Both $h=-\frac{8}{3} \lambda \Phi$ and
$\d_{\mu} h^{\mu \nu} = - \frac{2}{3} \lambda \d^{\nu} \Phi$ are
r-independent. Notice that $\Phi$ is essentially the brane-bending
\cite {gkr} degree of freedom. This can be seen from the
boundary conditions which are
\begin{equation}
 h_{\mu \nu}'=0 = h_{\mu \nu}^{TT'} + 2 e^{-2A} (\d_{\mu}
 \d_{\nu} - \Lambda g_{\mu \nu}) \Phi
\end{equation}
where TT stands for transverse traceless. In the absence of
matter, we can use the $\epsilon_5$
degree of freedom to
eliminate $\Phi$.

Although we introduced the $\Phi$ field in the previous analysis
to follow Refs. \cite{gregrub,quasipeople,kr}, we could have used
gauge invariance from the start to see that the introduction of
$\Phi$ was superfluous.
 A faster way to arrive at the same conclusion (that is
tracelessness of the graviton) is to first gauge fix the generic
solution to the $\mu$5 and 55 as much as possible before studying
the $\mu \nu$ equation. {}From the 55 equation, we get $e^{2A} h'=
c(x)$. The boundary conditions $e^{2A(0)} h'=0$ require $c(x)=0$.
So $h=f(x)$. As in a purely four dimensional
analysis, we then use the $\epsilon_{\mu}$ gauge
transformations to eliminate $f$. Similarly, the ${\mu 5}$
equations require $\d^{\mu} h_{\mu \nu}=g_{\nu}(x)$ and we can
eliminate the longitudinal parts as well. The remaining $h_{\mu
\nu}$ is now manifestly transverse and traceless. In both cases,
it is crucial that there is only a single brane; a second boundary
condition would ruin our ability to restrict to TT modes only.
For example, our analysis would not apply to the GRS model \cite{GRS}.

In the presence of sources, one could follow the procedure outlined
in \cite{gk}.  There, one first finds the profile for the trace and
longitudinal components of the metric.  These
no longer vanish but are given in terms of differential equations
involving $r$ only; that is, they do not propagate in the four-dimensional
spacetime.  One then uses these components as extra sources
to propagate the massive TT modes.
The upshot is that the only propagating degrees of freedom are the
massive TT modes. There is no ghost in the spectrum.

\section{Effective four-dimensional description}

An interesting question is if AdS$_{AdS}$ can be described
by an effective 4d action. Since the mass of the lowest mode is of
order $\Lambda^2$, it might be expected that the mass term arises
from terms that are quadratic in the curvature.
However it is easy to show that in four dimensions, local curvature
squared terms change neither the mass for the zero mode, nor the
$AdS$ curvature. That is, the background solution is unchanged, and
there is no mass generated for the linear perturbations. To see the
latter,
observe that
to linear order in the
fluctuation, and to second order in derivatives, the only gauge
invariant combination is the one set to zero by the linearized
Einstein equation, namely,
$$ R^{(1)}_{MN} -\frac{1}{2}g_{MN}R^{(1)} + 3\Lambda h_{MN}. $$
Here, $R^{(1)}_{MN}$ and $R^{(1)}$ are the linearized Ricci tensor
and Ricci scalar, respectively, while $h_{MN}$ is the metric
fluctuation. Thus, all linearized higher curvature terms are
either proportional to this combination, or are traces and
derivatives of it.  To find the possible mass terms to first order
in $L^2$, it is sufficient to substitute the solution to the
zeroth order Einstein equation into the above equation of motion.
As all higher curvature terms contain the above combination (set
to zero by Einstein's equations), one finds that such terms
vanish, and hence do not contribute a mass. This argument shows that
one cannot construct a local effective Lagrangian of metric
fluctuations alone that reproduces the physics of the
brane fluctuations.

Of course there are several other possibilities.
One way out would be to have a non-local 4d effective
action. It seems reasonable that this way one can produce
a mass term of order $\Lambda^2$ from higher curvature terms
by a mechanism analogous to the Debye mass in crystals.
This non-local effective action might be the result of
integrating out a CFT. In fact, it is clear that the 
construction of a 4D effective theory is subtle, since it is not
even true that all regions of space see
four-dimensional gravity.

\section{Absence of the VVDZ discontinuity}

As we emphasized in Section 2, the massive 4d graviton
always propagates 5 degrees of freedom. The absence
of the VVDZ discontinuity in AdS$_4$ was established in
\cite{mp,kogan} by a detailed study of the propagator.
Despite the difference in polarization states,
the massive propagator in AdS has a smooth limit as
the mass goes to zero. In the double scaling
limit, sending both the cosmological constant and the mass to zero,
modes whose $m^2$ go to zero faster than $\Lambda$ asymptote
to the massless graviton in 4d, while modes whose mass vanishes slower
asymptote to a massive graviton in 4d, which has a different propagator
from the massless graviton already at tree level, even when the mass
parameter is taken to zero.
It was questioned in \cite{duff}, whether this property
survives in the quantum theory.

In AdS$_{AdS}$ the mass squared of the
almost zero mode goes to zero as $\Lambda^2$, so
that it can smoothly go over into the massless graviton, while
the excited modes become the massive KK modes in the flat space limit.
From a 4d point of view, both the cosmological constant and the mass
will undergo renormalization. It is to be understood that
a large AdS$_4$ requires a fine tuning of
the quantum corrected cosmological constant.
The question now really becomes whether the mass of the almost
zero mode after quantum corrections still goes like $\Lambda^2$
and hence effectively reproduces massless physics, or
whether it looks more like one of the KK-modes, with $m^2$
of order $\Lambda$, which would signal a quantum VVDZ
discontinuity. From the 4d perspective, nothing seems
to protect the mass of the almost zero mode.

However from the 5d point of view, locality guarantees that
the low energy effective action is still of the same form
as the tree level action, with effective tension, Newton's
constant and bulk cc plus some higher derivative
corrections. The one fine tuning which we know we have to do
amounts to balancing effective tension versus effective
bulk cc. Once we have done this, the mass squared of the almost
zero mode will still go like $\Lambda_{eff}^2$, since
once we use the effective action, the full quantum answer is
given by the tree level analysis,
using the effective parameters instead of
the bare parameters is appropriated. Therefore there
is no VVDZ discontinuity even at the quantum level
in AdS$_{AdS}$.

{\bf Aknowlegements} We wish to thank Csaba Csaki, Josh Ehrlich,
Massimo Porrati, Matthew Schwartz,
Bob Wald, and Frank Wilczek for
useful conversations.

\end{multicols}
\end{document}